\documentclass[twocolumn,aps,superscriptaddress]{revtex4}

\usepackage{amssymb}
\usepackage{amsmath}
\usepackage{graphicx}
\usepackage[normalem]{ulem}
\usepackage[dvips]{color}
\usepackage{appendix}

\begin{document}

\title{Isospin effect on quark matter instabilities}
\author{Lu-Meng Liu}
\affiliation{School of Physical Sciences, University of Chinese Academy of Sciences, Beijing 100049, China}
\author{Wen-Hao Zhou}
\affiliation{Shanghai Institute of Applied Physics, Chinese Academy
of Sciences, Shanghai 201800, China}
\author{Jun Xu\footnote{Corresponding author: xujun@zjlab.org.cn}}
\affiliation{Shanghai Advanced Research Institute, Chinese Academy of Sciences, Shanghai 201210, China}
\affiliation{Shanghai Institute of Applied Physics, Chinese Academy of Sciences, Shanghai 201800, China}
\author{Guang-Xiong Peng}
\affiliation{School of Nuclear Science and Technology, University of Chinese Academy of Sciences, Beijing 100049, China}
\affiliation{Theoretical Physics Center for Science Facilities, Institute of High Energy Physics, Beijing 100049, China}
\affiliation{Synergetic Innovation Center for Quantum Effects $\&$ Applications, Hunan Normal University, Changsha 410081, China}
\date{\today}

\begin{abstract}
We have studied the mechanical and chemical instabilities as well as the liquid-gas-like phase transition in isospin asymmetric quark matter based on the NJL and the pNJL model. Areas of the mechanical instability region and the liquid-gas coexistence region are seen to be enlarged with a larger quark matter symmetry energy or in the presence of strange quarks. Our study shows that the light cluster yield ratio observed in relativistic heavy-ion collisions may not be affected much by the uncertainties of the isospin effect, and favors a smooth hadron-quark phase transition in compact stars as well as their mergers.
\end{abstract}

\maketitle


Extracting the phase diagram of Quantum Chromodynamics (QCD) from relativistic heavy-ion collisions and astrophysics helps us to understand properties of strong interactions. In the region of high temperatures or large baryon chemical potentials, reached in ultrarelativistic heavy-ion collisions or in the center of compact stars as well as their mergers, it is believed that the system is in the partonic phase with the chiral symmetry restored. In the region of low temperatures and small baryon chemical potentials, the system is in the hadronic phase with the chiral symmetry broken. Our knowledge on the transition from the hadronic phase to the partonic phase relies on lattice QCD calculations and effective QCD models. The former have told us that such phase transition at high temperatures and small baryon chemical potentials, reached in ultrarelativistic heavy-ion collisions, is a smooth crossover~\cite{Ber05,Aok06,Baz12a}. At large baryon chemical potentials, although the lattice QCD suffers from the sign problem, effective QCD models, such as the Nambu-Jona-Lasinio (NJL) model~\cite{Asa89,Fuk08,Car10,Bra13}, the Dyson-Schwinger approach~\cite{Xin14,Fis14}, and the functional renormalization group method~\cite{Fu19,Gao20}, tell us that the phase transition can be a first-order one. The QCD critical point along the phase boundary in-between the smooth crossover and the first-order phase transition is thus of crucial importance in mapping out the whole QCD phase diagram.

Aiming at searching for signals of the QCD critical point, 'low-energy' relativistic heavy-ion collisions have been carried out at RHIC-BES and NA61/SHINE and to be carried out at FAIR-CBM as well as NICA and HIAF, etc. If the trajectories of the system in these collisions pass through the critical point regions, one expects that exotic behaviors will occur. For instance, as a measure of the long-range correlation near the QCD critical point~\cite{Ste09,Asa09}, non-Gaussian net baryon fluctuations measured as the susceptibility of net proton fluctuations in RHIC-BES experiments show a non-monotonical behavior with respect to the collision energy~\cite{STAR10,STAR14,STAR20,STAR21a,STAR21b}. On the other hand, if the trajectories pass through the liquid-gas-like first-order phase transition region reached at even lower collision energies, one expects that the quark matter will be unstable and density fluctuations are expected to grow~\cite{Li16,Li17}. Recently, it was proposed from calculations based on the thermal and coalescence model that these density fluctuations may enhance the light cluster yield ratio $N_tN_p/N_d^2$~\cite{Sun17,Sun18}, with $N_p$, $N_d$, and $N_t$ being the number of protons, deuterons, and tritons, respectively. Based on the experimental data measured at SPS energies~\cite{Ant16} and RHIC-BES energies~\cite{Ada19,Zha19,Zha21}, the non-monotonic behavior of the light cluster yield ratio with respect to the collision energy has been observed. More recently, starting from a thermalized parton distribution, it has been illustrated that the light cluster yield ratio does increase using the NJL transport model with a first-order phase transition, compared to that with a smooth crossover~\cite{Sun20}.

Observables of compact stars provide possibilities of studying the QCD phase diagram at extremely high densities~\cite{Hem13}. The gravitational wave emitted from compact star mergers may carry information of the hadron-quark phase transition~\cite{Mos19,Bau19,Wei20,Bau20} in the formed matter at a few times of normal nuclear matter density and the temperature around 100 MeV. It has been argued that such phase transition in the so-called hybrid stars~\cite{Hem09,Alf13} or compact star mergers~\cite{Mos19,Bau19} could be a first-order one, and this has led to further theoretical studies based on the Bayesian analysis~\cite{Bla20a,Tan20,Xie21}. A first-order phase-transition of strong interacting matter in astrophysics systems may result in the softness of the equation of state as well as exotic structures in the hadron-quark mixed phase~\cite{Hei93,Mar07,Bla20}.

Although studying light cluster yield ratios in RHIC and compact star mergers are useful in searching for the spinodal region of the QCD phase diagram, a systematic study of the isospin effect on the phase transition and instabilities in the strong interacting quark matter in these systems is still lacking~\cite{Cos20}. It is known that heavy-ion collisions induced by neutron-rich nuclei form isospin asymmetric systems, and the isospin asymmetry is expected to be increasingly important in collisions at lower collision energies with a larger stopping power and a larger baryon chemical potential, and nerveless to say, in the center of compact stars or their mergers where large isospin asymmetries are reached. In isospin asymmetric nuclear matter formed by unequal numbers of neutrons and protons, it has been shown that the properties of the liquid-gas phase transition as well as the mechanical and chemical instabilities are largely different from those in isospin symmetric nuclear matter (see, e.g., Ref.~\cite{Li08} and references therein), and the isospin effect~\cite{Xu08} depends on the nuclear symmetry energy characterizing the energy difference between isospin asymmetric and symmetric matter. In the isospin asymmetric quark matter formed in low-energy relativistic heavy-ion collisions or in compact star mergers, it is expected that the isospin effect or the quark matter symmetry energy will affect the properties of the liquid-gas-like phase transition and the instabilities of quark matter, and consequently observables from RHIC and astrophysics mentioned above.



In this manuscript, we study the isospin effect on the quark matter instabilities based on the 3-flavor NJL model as well as its Polyakov-loop extension. We start from the Lagrangian of the 3-flavor NJL model written as
\begin{eqnarray}
\mathcal{L}_{\rm NJL} &=& \bar{\psi}(i\rlap{\slash}\partial-\hat{m})\psi
+\frac{G_S}{2}\sum_{a=0}^{8}[(\bar{\psi}\lambda_a\psi)^2+(\bar{\psi}i\gamma_5\lambda_a\psi)^2]
\notag\\
&-&\frac{G_V}{2}\sum_{a=0}^{8}[(\bar{\psi}\gamma_\mu\lambda_a\psi)^2+
(\bar{\psi}\gamma_5\gamma_\mu\lambda_a\psi)^2]
\notag\\
&-&G_{IV}\sum_{a=1}^{3}[(\bar{\psi}\gamma_\mu\lambda_a\psi)^2+(\bar{\psi}\gamma_5\gamma_\mu\lambda_a\psi)^2]\notag\\
&-&K\{\det[\bar{\psi}(1+\gamma_5)\psi]+\det[\bar{\psi}(1-\gamma_5)\psi]\}.
\end{eqnarray}
In the above, $\psi = (u, d, s)^T$ and $\hat{m} = \text{diag}(m_u, m_d, m_s)$ are the quark fields and the current quark mass matrix for $u$, $d$, and $s$ quarks, respectively; $\lambda_a$ are the Gell-Mann matrices with $\lambda_0$ = $\sqrt{2/3}I$ in the 3-flavor space with the SU(3) symmetry; $G_S$, $G_V$, and $G_{IV}$ are respectively the scalar-isoscalar, vector-isoscalar, and vector-isovector coupling constant; and the $K$ term represents the six-point Kobayashi-Maskawa-t' Hooft interaction that breaks the axial $U(1)_A$ symmetry. In the present study, we adopt the parameters $m_u = m_d = 3.6$ MeV, $m_s = 87$ MeV, $G_S\Lambda^2 = 3.6$, $K\Lambda^5 = 8.9$, and the cutoff value in the momentum integral $\Lambda = 750$ MeV/c given in Refs.~\cite{Bra13,Lut92}. The position of the critical point is sensitive to $G_V$~\cite{Asa89,Fuk08,Bra13}, characterizing the QCD phase diagram and the quark matter equation of state (EOS) at large baryon chemical potentials. In the present study, we neglect the vector-isoscalar contribution by setting $G_V=0$ while focusing on the isospin effect on the quark matter instabilities, i.e., properties of isospin asymmetric quark matter with a finite reduced vector-isovector coupling constant $R_{IV}=G_{IV}/G_S$.

From the mean-field approximation, the thermodynamic potential
$\Omega_\textrm{NJL}$ of quark matter from the finite-temperature field theory can be expressed as
\begin{eqnarray}\label{omeganjl}
\Omega_{\textrm{NJL}}&=& -2N_c\sum_{q=u,d,s}\int_0^\Lambda\frac{d^3p}{(2\pi)^3}
[E_q+T\ln(1+e^{-\beta(E_q-\tilde{\mu}_q)})
\notag\\
&+&T\ln(1+e^{-\beta(E_q+\tilde{\mu}_q)})]+G_S(\phi_u^2+\phi_d^2+\phi_s^2)
\notag\\
&-&4K\phi_u\phi_d\phi_s
-G_{IV}(\rho_u-\rho_d)^2.
\end{eqnarray}
In the above, the factor $2N_c=6$ represents the spin and color degeneracy of the quark, $\beta=1/T$ is the inverse of the temperature, $\phi_q$ is the quark condensate for quark flavor $q$ expressed as
\begin{equation}\label{sigmaq}
\phi_q=-2N_c\int_0^{\Lambda}\frac{d^3p}{(2\pi)^3}\frac{M_q}{E_q}(1-f_q-\bar{f_q}),
\end{equation}
where $E_q=\sqrt{p^2 +M_q^2}$ is the single-quark energy, and the in-medium Dirac mass $M_q$ is related to the quark condensate through the relations
\begin{eqnarray}
M_{u} &=& m_{u}-2G_S\phi_{u}+2K\phi_{d}\phi_{s}, \label{mass1}\\
M_{d} &=& m_{d}-2G_S\phi_{d}+2K\phi_{s}\phi_{u}, \label{mass2}\\
M_{s} &=& m_{s}-2G_S\phi_{s}+2K\phi_{u}\phi_{d}. \label{mass3}
\end{eqnarray}
The phase-space distribution functions $f_q$ and $\bar{f_q}$ for quarks and antiquarks in Eq.~(\ref{sigmaq}) can be expressed as the Fermi-Dirac distributions in the thermodynamical calculation, i.e.,
\begin{eqnarray}
f_q &=& \frac{1}{\exp[(E_q-\tilde{\mu}_q)/T]+1},\label{fqnjl}\\
\bar{f_q} &=& \frac{1}{\exp[(E_q+\tilde{\mu}_q)/T]+1},\label{fqbnjl}
\end{eqnarray}
where $\tilde{\mu}_q$ is the effective chemical potential which is related to the real chemical potential $\mu_q$ through the relations
\begin{equation}\label{muq}
\tilde{\mu}_q = \mu_q
-2G_{IV}\tau_q(\rho_u-\rho_d), \\
\end{equation}
with $\tau_u=1$, $\tau_d=-1$, and $\tau_s=0$. In the above, $\rho=\rho_u+\rho_d+\rho_s$ is the total net quark density, and for a single quark flavor $q$ the net quark density $\rho_q$ can be calculated from
\begin{equation}\label{rhoqnjl}
\rho_q=2N_c\int^{\Lambda}_0(f_q-\bar{f_q})\frac{d^3p}{(2\pi)^3}.
\end{equation}
Since the quark condensates and the Dirac masses are related to each other, Eqs.~(\ref{sigmaq}) and (\ref{mass1}-\ref{mass3}) can be solved self-consistently through the iteration method. The quark condensate or the Dirac mass is the order parameter of the chiral phase transition.

The thermodynamic potential $\Omega_\textrm{pNJL}$ of the 3-flavor
pNJL model at finite temperature and quark chemical potential can be
expressed as
\begin{eqnarray}\label{omegapnjl}
\Omega_{\textrm{pNJL}} &=&\mathcal{U}(\Phi,\bar{\Phi},T)-2N_c\sum_{q=u,d,s}\int_0^\Lambda\frac{d^3p}{(2\pi)^3}E_q
\notag\\
&-&2T\sum_{q=u,d,s}\int\frac{d^3p}{(2\pi)^3}[\ln(1+e^{-3\beta(E_q-\tilde{\mu}_q)}
\notag\\
&+&3\Phi e^{-\beta(E_q-\tilde{\mu}_q)}
+3\bar{\Phi}e^{-2\beta(E_q-\tilde{\mu}_q)})
\notag\\
&+&\ln(1+e^{-3\beta(E_q+\tilde{\mu}_q)}
+3\bar{\Phi} e^{-\beta(E_q+\tilde{\mu}_q)}
\notag\\
&+&3\Phi e^{-2\beta(E_q+\tilde{\mu}_q)})]
+G_S(\phi_u^2+\phi_d^2+\phi_s^2)
\notag\\
&-&4K\phi_u\phi_d\phi_s
-G_{IV}(\rho_u-\rho_d)^2.
\end{eqnarray}
Reference~\cite{Fuk04} first introduced the Polyakov loop contribution, with $\Phi$ ($\bar{\Phi}$) related to the excess free
energy for a static quark (anti-quark) in a hot gluon medium~\cite{Fuk11}, and thus serving as an order parameter for the deconfinement phase transition. The form of the temperature-dependent effective potential
$\mathcal{U}(\Phi, \bar{\Phi}, T)$ as a function of the Polyakov loop $\Phi$ and $\bar{\Phi}$ in Eq.~(\ref{omegapnjl}) is taken from Ref.~\cite{Fuk08}, i.e.,
\begin{eqnarray}
\mathcal{U}(\Phi,\bar{\Phi},T) &=& -b \cdot
T\{54e^{-a/T}\Phi\bar{\Phi} +\ln[1-6\Phi\bar{\Phi}
\notag\\
&-&3(\Phi\bar{\Phi})^2+4(\Phi^3+\bar{\Phi}^3)]\}.
\end{eqnarray}
The parameters
$a=664$ MeV and $b=0.028\Lambda^3$ are determined by the condition
that the first-order phase transition in the pure gluodynamics takes
place at $T = 270$ MeV~\cite{Fuk08}, and the simultaneous crossover of the chiral
restoration and the deconfinement phase transition occurs around $T
\approx 212$ MeV. In order to get the minimum of the thermodynamic potential
$\Omega_\textrm{pNJL}$, the following five equations are solved
\begin{eqnarray}
\frac{\partial\Omega_{\textrm{pNJL}}}{\partial\phi_u}
=\frac{\partial\Omega_{\textrm{pNJL}}}{\partial\phi_d}
=\frac{\partial\Omega_{\textrm{pNJL}}}{\partial\phi_s}
=\frac{\partial\Omega_{\textrm{pNJL}}}{\partial\Phi}
=\frac{\partial\Omega_{\textrm{pNJL}}}{\partial\bar{\Phi}} =0,
\notag
\end{eqnarray}
leading to the values of $\phi_u$, $\phi_d$, $\phi_s$, $\Phi$,
$\bar{\Phi}$ and also other thermodynamical quantities in the pNJL model.

Starting from the thermodynamic potential, the energy density of the
system can be obtained from the thermodynamical relation
\begin{equation}
\varepsilon=\Omega+\beta\frac{\partial}{\partial\beta}\Omega+\sum_{q=u,d,s}\mu_q\rho_q.
\end{equation}
Accordingly, the energy density from the NJL model can
be written as
\begin{eqnarray}\label{epsilon}
\varepsilon_{\textrm{NJL}}&=&-2N_c\sum_{q=u,d,s}\int_0^{\Lambda}\frac{d^3p}{(2\pi)^3}
E_q(1-f_q-\bar{f_q})
\notag\\
&+&G_S(\phi_u^2+\phi_d^2+\phi_s^2)-4K\phi_u\phi_d\phi_s\notag\\
&+&G_{IV}(\rho_u-\rho_d)^2
- \varepsilon_0.
\end{eqnarray}
In the above expression, $\varepsilon_0$ is the quark condensate contribution in vacuum, and thus ensures
$\varepsilon_{\textrm{NJL}} = 0$ in vacuum. Similarly, the energy
density from the pNJL model can be expressed as
\begin{eqnarray}\label{epsilon_pNJL}
\varepsilon_{\textrm{pNJL}}&=&
54abe^{-a/T}\Phi\bar{\Phi}-2N_c\sum_{q=u,d,s}\int^\Lambda_0\frac{d^3p}{(2\pi)^3}
E_q
\notag\\
&+&2N_c\sum_{q=u,d,s}\int\frac{d^3p}{(2\pi)^3}
E_q(F_q+\bar{F_q})\notag\\
&+&G_S(\phi_u^2+\phi_d^2+\phi_s^2)-4K\phi_u\phi_d\phi_s\notag\\
&+&G_{IV}(\rho_u-\rho_d)^2
-\varepsilon_0,
\end{eqnarray}
where $F_q$ and $\bar{F_q}$ are the effective phase-space distribution functions for quarks and
antiquarks in the pNJL model, and they can be expressed as
\begin{equation}\label{Fq}
F_q = \frac{1+2\bar\Phi\xi_q+\Phi\xi_q^2}{1+ 3\bar\Phi\xi_q
+3\Phi\xi_q^2+\xi_q^3}
\end{equation}
and
\begin{equation}\label{Fqb}
\bar{F_q}=
\frac{1+2\Phi{\xi^\prime_q}+\bar\Phi{\xi^\prime_q}^2}{1+3\Phi{\xi^\prime_q}+3\bar\Phi{\xi^\prime_q}^2+{\xi^\prime_q}^3}
\end{equation}
with $\xi_q =
e^{(E_q-\tilde{\mu}_q)/T}$ and $\xi_q^\prime =
e^{(E_q+\tilde{\mu}_q)/T}$. Similarly, the net quark density in the pNJL model is related to the effective phase-space distribution functions through the relation
\begin{equation}
\rho_q=2N_c\int(F_q-\bar{F_q})\frac{d^3p}{(2\pi)^3}.
\end{equation}

For both the NJL and the pNJL model, the pressure $P$ can be calculated from the thermodynamic potential
\begin{equation}
P = -\Omega + \Omega_0,
\end{equation}
where $\Omega_0=\varepsilon_0$ is the thermodynamical potential in vacuum, to ensure the pressure is zero in vacuum. Note that the pNJL model reduces to the NJL model at zero temperature, and we will compare their behaviors at finite temperatures.


\begin{figure}[ht]
	\includegraphics[scale=0.35]{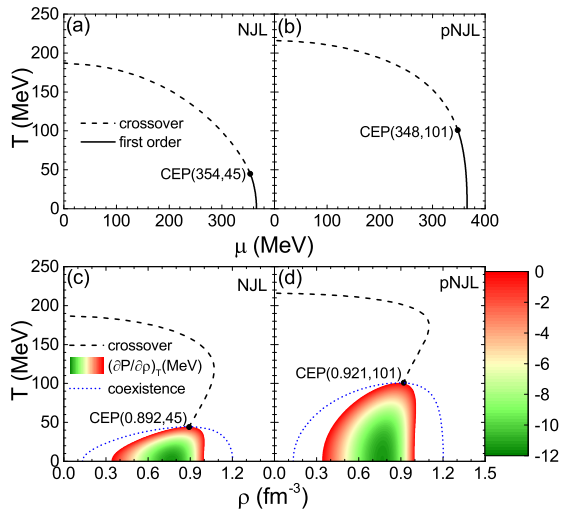}
	\caption{(Color online) Phase diagrams from the NJL (left) and the pNJL (right) model for isospin symmetric quark matter. Upper: Phase diagrams in the $(T,\mu)$ plane showing the chiral phase transition. Lower: Phase diagrams in the $(T,\rho)$ plane showing regions of the liquid-gas phase coexistence and the instability. $\mu=\mu_u=\mu_d$ is the quark chemical potential and $\rho$ is the net quark density. $\mu_s$ is set to 0 and $R_{IV}$ is set to 0 in this plot.}\label{fig1}
\end{figure}

We first display the phase diagrams from the NJL and the pNJL model for isospin symmetric quark matter in the $(T,\mu)$ plane and $(T,\rho)$ plane in Fig.~\ref{fig1}. The dashed lines in both upper and lower panels indicate the smooth change of the quark condensate or the Dirac quark mass, and they are thus a smooth crossover. The solid lines in the upper panels show the phase boundary of the first-order chiral phase transition in the $(T,\mu)$ plane, numerically corresponding to a sudden drop of the Dirac quark mass with increasing $\mu$ for a given $T$. Starting from the same thermodynamic potential, the (p)NJL model is able to describe both the chiral phase transition and the liquid-gas-like phase transition, and they are related to each other, i.e., in the region of the first-order chiral phase transition the pressure $P$ and the chemical potential $\mu$ change non-monotonically with increasing $\rho$, leading to instabilities and the liquid-gas-like phase transition. The chemical potential of the first-order chiral phase transition boundary in the $(T,\mu)$ plane can be regarded as that in the liquid-gas mixed phase, and the boundaries of the liquid-gas phase coexistence region are plotted in the lower panels of Fig.~\ref{fig1}. In addition, the mechanical instability regions ($(\partial P/\partial \rho)_T<0$) are also displayed in the lower panels of Fig.~\ref{fig1}, with a smaller area compared with the phase coexistence region. The temperature above which there is no instability region is exactly that of the critical end point (CEP), in-between the phase boundaries of the smooth crossover and the first-order phase transition, as shown in the phase diagrams in both the $(T,\mu)$ plane and the $(T,\rho)$ plane. Due to the different phase-space distribution functions, the pNJL model has the CEP at a higher temperature and has a larger instability region, compared with the NJL model.

\begin{figure}[ht]
	\includegraphics[scale=0.3]{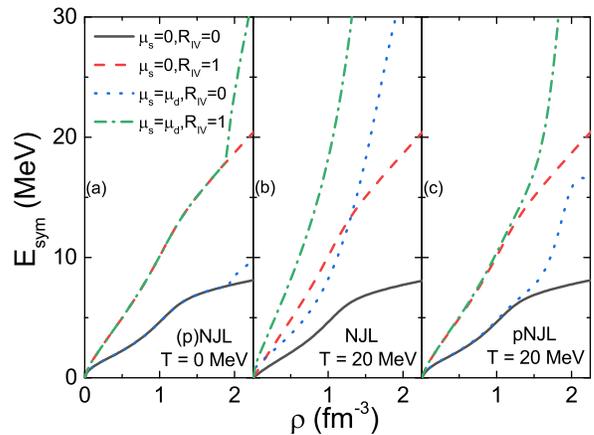}
	\caption{(Color online) Quark matter symmetry energy as a function of net quark density from the NJL and pNJL model for cold ($T=0$ MeV) and hot ($T=20$ MeV) quark matter, with $\mu_s=0$ and $\mu_s=\mu_d$ as well as $R_{IV}=0$ and $R_{IV}=1$.}\label{fig2}
\end{figure}

Figure~\ref{fig1} shows the phase diagram in a typical case with zero net strange quark density and without isovector coupling. In the present study, we also consider the case with $s$ quarks being in chemical equilibrium with $d$ quarks, i.e., $\mu_s=\mu_d$, as well as a strong isovector coupling $R_{IV}=1$ favored by the elliptic flow splitting between $\pi^-$ and $\pi^+$ in relativistic heavy-ion collisions at RHIC-BES energies~\cite{Liu19}. The purpose is to investigate effects of strangeness and isovector coupling on the instability properties of isospin asymmetric quark matter formed at RHIC or in compact stars, whose EOS is largely characterized by the quark matter symmetry energy defined as
\begin{equation}
E_{sym}(\rho,\mu_s) = \frac{1}{2}\frac{\partial^2E(\rho, \delta, \mu_s)}{\partial\delta^2}\mid_{\delta=0},
\end{equation}
where $E=\epsilon/\rho$ is the energy per particle, and $\delta=3(\rho_d-\rho_u)/(\rho_d+\rho_u)$ is the isospin asymmetry. A similar definition of the nuclear matter symmetry energy and its effect on the instability properties of isospin asymmetric nuclear matter can be found in Ref.~\cite{Xu08}. Figure~\ref{fig2} compares the quark matter symmetry energy as a function of net quark density $\rho$ in different scenarios. It is seen that $s$ quarks affect $E_{sym}$ only after they appear at very high $\rho$ at zero temperature, while their effects become considerably large at finite temperatures when there is no such threshold effect for $s$ quarks, especially for the NJL model. In the case of $\mu_s=\mu_d$, the contribution of $s$ quarks to the total EOS increases with increasing $\delta$, thus their appearance always increase $E_{sym}$. This is different from the case of fixing the net $s$ quark density as shown in Ref.~\cite{Liu16}. Moreover, it is seen that a stronger isovector coupling increases $E_{sym}$ and stiffens the EOS of isospin asymmetric quark matter.

\begin{figure}[ht]
	\includegraphics[scale=0.35]{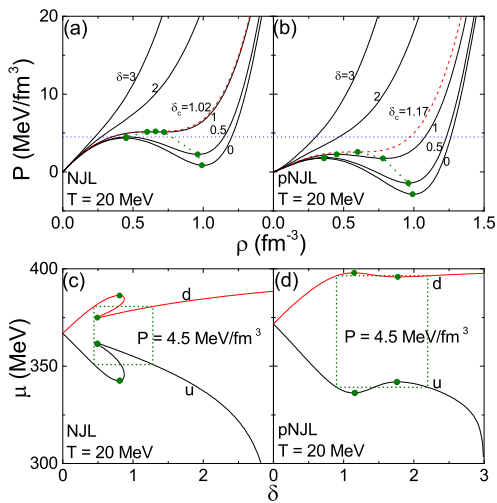}
	\caption{(Color online) Pressure as a function of net quark density (upper) and the $u$ and $d$ quark chemical potential isobars at a fixed pressure $P=4.5$ MeV/fm$^3$ as a function of the isospin asymmetry $\delta$ (lower) from the NJL (left) and the pNJL (right) model for hot quark matter at $T=20$ MeV. Rectangles representing the Maxwell construction are plotted in lower panels to identify the liquid and gas phase that can coexist. $\mu_s$ is set to 0 and $R_{IV}$ is set to 0 in this plot.}\label{fig3}
\end{figure}

We illustrate how we construct the mechanical and chemical instability regions in Fig.~\ref{fig3} for a typical case with $\mu_s=0$ and $R_{IV}=0$. The upper panels of Fig.~\ref{fig3} show how the pressures in quark matter of different isospin asymmetries change with the net quark density at a given temperature $T=20$ MeV. The mechanical instability regions shown in the lower panels of Fig.~\ref{fig1} correspond to those with $(\partial P/\partial \rho)_T<0$. In such regions, an infinitely small increase of $\rho$ will further reduce the pressure, and the quark matter will be further compressed and the density will be even higher. Thus, a small fluctuation of $\rho$ makes the system mechanically unstable. It is seen that once the isospin asymmetry increases to $\delta_c=1.02$ for the NJL model and $\delta_c=1.17$ for the pNJL model at $T=20$ MeV, the pressure increases monotonically with increasing $\rho$, and there is no mechanical instability region. The lower panels of Fig.~\ref{fig3} show how the chemical potentials of $u$ and $d$ quarks evolve with the isospin asymmetry at a fixed pressure $P=4.5$ MeV/fm$^3$. In the region of $(\partial \mu_d/\partial \delta)_P<0$ or $(\partial \mu_u/\partial \delta)_P>0$, an infinitely small increase of $\delta$ will reduce (increase) the chemical potential of $d$ ($u$) quarks, and more $d$ ($u$) quarks will enter (leave) the quark matter and the isospin asymmetry will be further increased. Thus, a small fluctuation of $\delta$ makes the system chemically unstable.

\begin{figure}[ht]
	\includegraphics[scale=0.45]{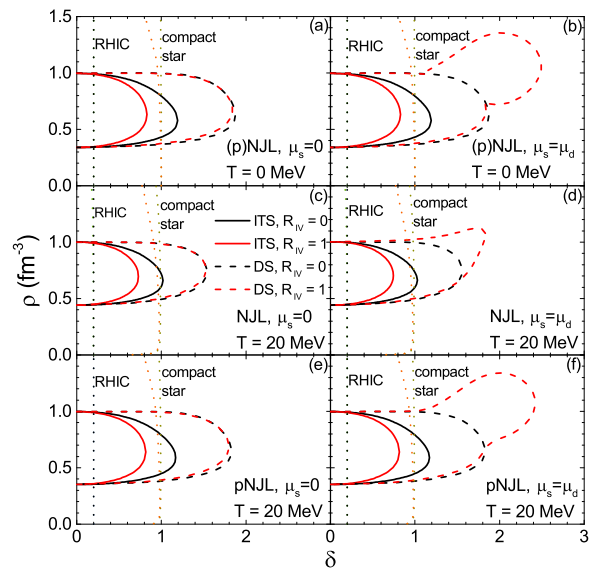}
	\caption{(Color online) Mechanical and chemical instability regions in the $(\rho,\delta)$ plane from the NJL and the pNJL model for cold ($T=0$) and hot ($T=20$ MeV) quark matter, with $\mu_s=0$ (left) and $\mu_s=\mu_d$ (right) as well as $R_{IV}=0$ and $R_{IV}=1$.}\label{fig4}
\end{figure}

The isothermal spinodal (ITS) and the diffusive spinoal (DS) boundaries representing respectively those for the mechanical and chemical instability regions in the $(\rho,\delta)$ plane for different scenarios are compared in Fig.~\ref{fig4}. The dotted lines around $\delta=0.2$ are from the condition that the net-charge density is $40\%$ of the total net-baryon density, corresponding to the case of relativistic heavy-ion collisions (RHIC), while those around $\delta=1$ (slightly different for $R_{IV}=0$ and $R_{IV}=1$) are from the charge neutrality and $\beta$-equilibrium condition, corresponding to the case of compact stars. The areas on the left-hand side of the solid lines are the mechanical instability regions where the system is chemically stable, while those surrounded by both the solid lines and the dashed lines are the chemical instability regions where the system is mechanically stable. In both areas, the system is unstable, but how the fluctuation grows with $\rho$ or $\delta$ may depend on the detailed instability properties, and this can be a future study. It is seen that in the RHIC case at about $\delta=0.2$ the mechanical instability is adequate to describe the instability of the system, while in the compact star case around $\delta=1$, one has to consider the chemical instability, and the detailed instability properties depend on different scenarios. It is seen that a stronger isovector coupling leading to a larger quark matter symmetry energy generally stiffens the EOS of isospin asymmetric quark matter, and thus reduces the area of the mechanically instability region. The boundary of the chemical instability region is not much affected by the value of $R_{IV}$ for $\mu_s=0$, while it can be further expanded to areas of larger $\delta$ for $\mu_s=\mu_d$, where $s$ quarks appear. The area of the instability region is generally reduced with increasing temperature, while the temperature effect is seen to be weaker for the pNJL model compared with the NJL model.

\begin{figure}[ht]
	\includegraphics[scale=0.3]{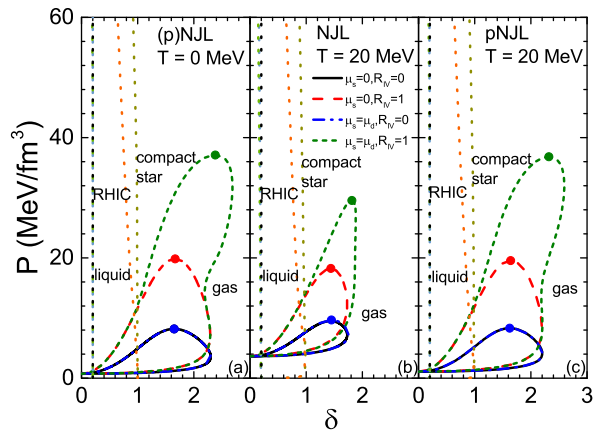}
	\caption{(Color online) Binodal surface in the $(\rho,\delta)$ plane from the NJL and the pNJL model for cold ($T=0$) and hot ($T=20$ MeV) quark matter, with $\mu_s=0$ and $\mu_s=\mu_d$ as well as $R_{IV}=0$ and $R_{IV}=1$. }\label{fig5}
\end{figure}

The liquid-gas-like phase transition of asymmetric quark matter can be studied by doing the Maxwell construction~\cite{Mul95,Xu08}, i.e., by plotting the rectangle in the chemical potential isobar as in the lower panels of Fig.~\ref{fig3}. In this way, one finds the two states satisfying the Gibbs condition, i.e., they have the same quark chemical potentials, the same pressure, and the same temperature, and can thus coexist. The liquid (gas) phase has a higher (lower) density and a smaller (larger) isospin asymmetry. Varying the pressure and collecting all the liquid and gas states leads to the binodal surface, as shown in Fig.~\ref{fig5} for various scenarios. The small-$\delta$ (large-$\delta$) side of the binodal surface corresponds to the liquid (gas) phase, while within the binodal surface there can only exist the liquid-gas mixed phase. There is a maximum pressure above which there is no chemical instability region and thus no liquid-gas mixed phase. It is seen that a stronger isovector coupling corresponding to a larger quark matter symmetry energy leads to a larger area of the liquid-gas mixed phase region, consistent with the case of nuclear matter~\cite{Xu08}. On the other hand, although the results from $\mu_s=\mu_d$ are similar to those from $\mu_s=0$ for $R_{IV}=0$, one observes a higher maximum pressure and a larger liquid-gas coexistence region from $\mu_s=\mu_d$ compared with $\mu_s=0$, due to the appearance of $s$ quarks, in the presence of a strong isovector coupling. The phase coexistence region generally shrinks with the increasing temperature, while the temperature effect is weaker for the pNJL model compared with the NJL model. Again, the dotted lines for the RHIC case around $\delta=0.2$ and those for the compact star case around $\delta=1$ are also shown for comparison. It is seen that different scenarios leads to similar coexistence regions for the RHIC case, while for the compact star case the liquid-gas mixed phase region depends on the isovector coupling, the existence of $s$ quarks, etc.

\begin{figure}[ht]
	\includegraphics[scale=0.35]{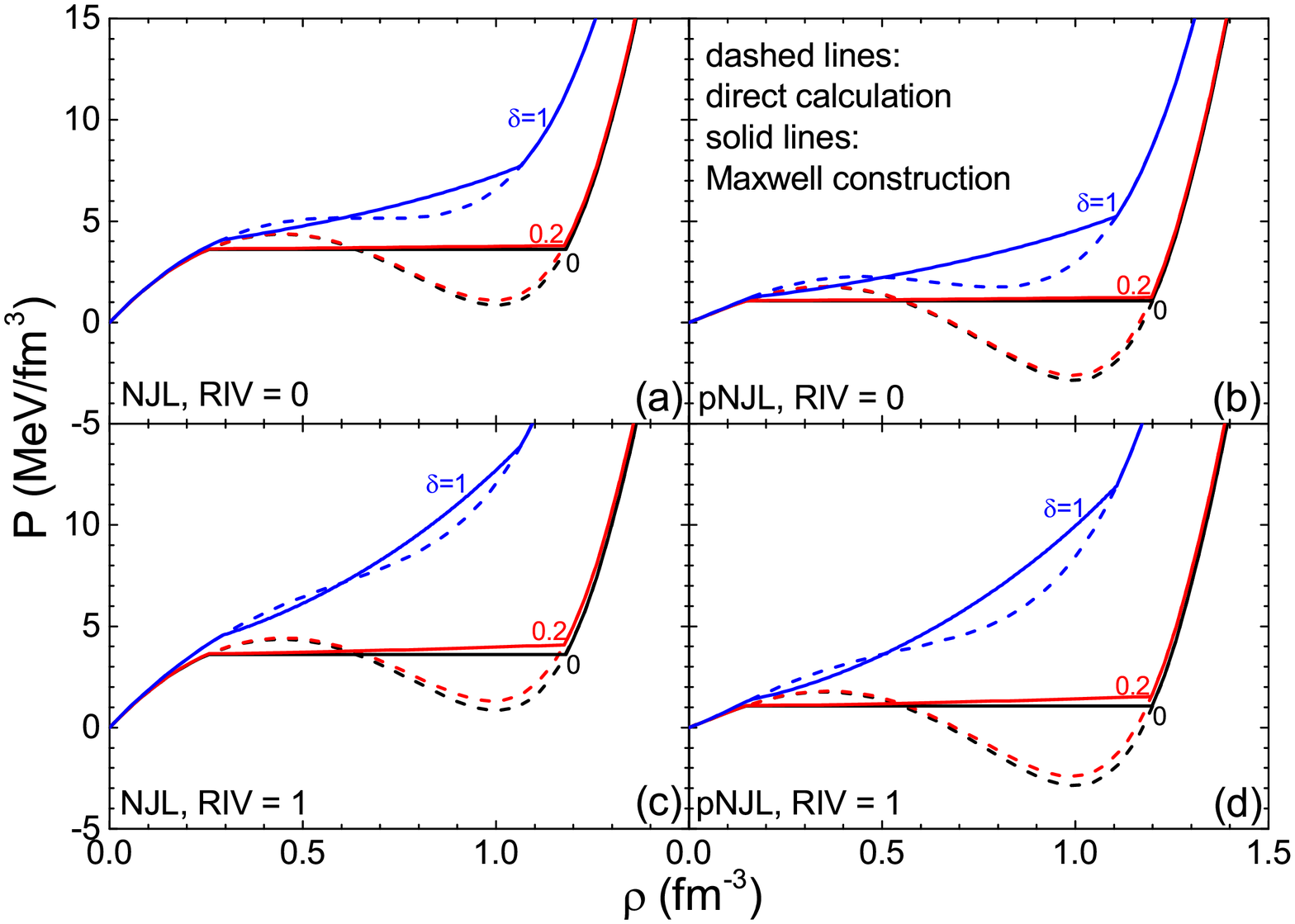}
	\caption{(Color online) Pressure as a function of net quark density for quark matter of different isospin asymmetries at a fixed temperature $T=20$ MeV showing the liquid-gas-like phase transition. $\mu_s$ is set to 0 and $R_{IV}$ is set to 0 (upper) and 1 (lower).}\label{fig6}
\end{figure}

\begin{figure}[ht]
	\includegraphics[scale=0.35]{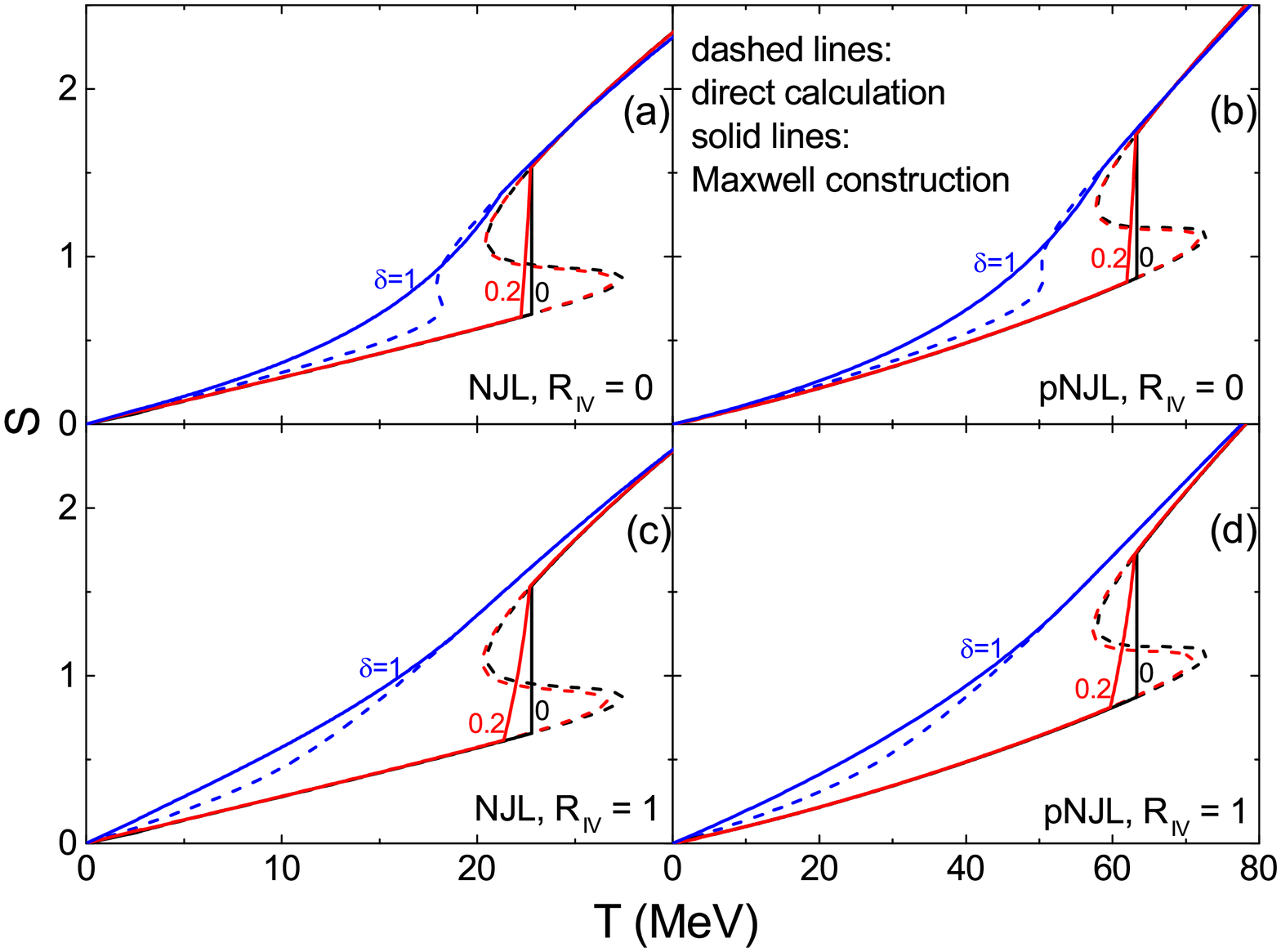}
	\caption{(Color online) Entropy per particle as a function of temperature for quark matter of different isospin asymmetries at a fixed pressure $P=4.5$ MeV/fm$^3$ showing the liquid-gas-like phase transition. $\mu_s$ is set to 0 and $R_{IV}$ is set to 0 (upper) and 1 (lower).}\label{fig7}
\end{figure}

The binodal surface shown in Fig.~\ref{fig5} is useful in constructing the liquid-phase transition in asymmetric quark matter. For details of the analysis method, we refer the reader to Refs.~\cite{Xu08}. Figure~\ref{fig6} displays how the pressure evolves with the increasing density in the presence of the liquid-gas-like phase transition at a fixed temperature $T=20$ MeV for different scenarios. The dashed lines are from direct calculations by assuming that the quark matter remains in one phase, while the corresponding solid lines in the phase transition region are from the coexistence of the liquid phase and the gas phase, with $P$ being the same pressure in both phases but $\rho$ being the average density. It is seen that for isospin symmetric quark matter ($\delta=0$) the pressure remains unchanged during the liquid-gas-like phase transition, showing a first-order phase transition, and is exactly corresponding to the phase coexistence region in the lower panels of Fig.~\ref{fig1}. In the RHIC case with $\delta=0.2$ the phase transition can be regarded approximately as the first-order one. In the compact star case with $\delta=1$ the pressure increases smoothly with the increasing density, corresponding to a smooth crossover. A stronger isovector coupling stiffens the EOS, while the pNJL model has a softer EOS compared to the NJL model. Instead of compressing the asymmetric quark matter at a given temperature, it is also possible to study the liquid-gas-like phase transition by heating the asymmetric quark matter at a fixed external pressure. Figure~\ref{fig7} displays how the entropy per particle evolves with the increasing temperature in the later case at $P=4.5$ MeV/fm$^3$ for different scenarios, where the entropy per particle $S$ is calculated from
\begin{eqnarray}
S &=& -\frac{2N_c}{\rho} \sum_{q=u,d,s} \int\frac{\mathrm{d}^3p}{\left(2\pi\right)^3}\left[n_q \ln  n_q+\left(1-n_q\right)\ln\left(1-n_q\right)\right] \notag\\
 &-& \frac{2N_c}{\rho} \sum_{q=u,d,s} \int\frac{\mathrm{d}^3p}{\left(2\pi\right)^3}\left[\bar{n}_q \ln  \bar{n}_q +\left(1-\bar{n}_q \right)\ln\left(1-\bar{n}_q \right)\right], \notag\\\label{sden}
\end{eqnarray}
with $n_q=f_q$ and $\bar{n}_q=\bar{f_q}$ for the NJL model and $n_q=F_q$ and $\bar{n}_q=\bar{F_q}$ for the pNJL model, respectively. Again, the dashed lines are from direct calculations, while the solid lines in the corresponding phase coexistence region are from averaging the entropy density in the liquid phase and the gas phase with the same temperature $T$. It is interesting to see that the temperature remains unchanged during the liquid-gas-like phase transition for isospin symmetric quark matter ($\delta=0$). In the RHIC case with $\delta=0.2$ such phase transition becomes much smoother, especially for $R_{IV}=1$ compared with $R_{IV}=0$, while in the compact star case with $\delta=1$ the phase transition becomes a smooth crossover. The phase transition temperatures for $\delta=0$ and $\delta=0.2$ are much higher in the pNJL model compared to those in the NJL model. Both Figs.~\ref{fig6} and \ref{fig7} are for $\mu_s=0$, while it is found that the results are similar to those for $\mu_s=\mu_d$ at the temperatures and pressures considered here.

To summarize, based on the NJL and the pNJL model, we have studied the instabilities and liquid-gas-like phase transition in quark matter formed in relativistic heavy-ion collisions (RHIC) or in compact stars as well as their mergers. Such instabilities are expected to affect the light cluster yield ratios in RHIC proposed as a probe of mapping out the QCD phase diagram, or the equation of state or exotic structures in the hadron-quark mixed phase in astrophysics systems. Although the isospin effect on the instability region reached at RHIC is small due to the small isospin asymmetry, it is found that the liquid-gas-like phase transition becomes a smooth one in isospin asymmetric quark matter, compared with the first-order phase transition in isospin symmetric quark matter. In compact stars or their mergers with large isospin asymmetries, areas of the mechanical instability and the phase coexistence region increase with increasing strength of the isovector coupling and strangeness degree of freedom in the quark matter system, and the hadron-quark phase transition is likely to be a smooth crossover.



JX was supported by the National Natural Science Foundation of China under Grant No. 11922514. GXP was supported by the National Natural Science Foundation of China under Grant Nos. 11875052, 11575190, and 11135011.


\end{document}